\documentclass[a4paper]{jpconf}
\usepackage{graphicx}
\usepackage{epstopdf}
\begin{document}
\title{Overview on jet results from STAR}

\author{Elena Bruna, for the STAR Collaboration}

\address{Physics Department, Yale University \\ 
272 Whitney Avenue, New Haven, CT-06511 USA}

\ead{elena.bruna@yale.edu}

\begin{abstract}
Full jet reconstruction allows access to the parton kinematics over a large energy domain and can be used to constrain the mechanisms of energy loss in heavy-ion collisions.
Such measurements are challenging at RHIC, due to the high-multiplicity environments created in heavy-ion collisions. 
In these proceedings, we report an overview of the results on full jet reconstruction obtained by the STAR experiment. Jet measurements in 200 GeV p+p show that jets are calibrated pQCD probes and provide a baseline for jet measurements in Au+Au collisions.
Inclusive differential jet production cross sections and ratios are reported for central 200 GeV Au+Au collisions and compared to p+p. We also present measurements of fully reconstructed di-jets at mid-rapidity, and compare spectra and fragmentation functions in p+p and central Au+Au collisions.
\end{abstract}

\section{Introduction}
Jet quenching at RHIC was observed via single hadron and di-hadron suppression  at high $p_T$ in Au+Au collisions relative to p+p and d+Au~\cite{Raa}.
Although these measurements showed evidence for parton energy loss in the medium, they were limited by the inability to constrain the kinematics of the hard-scattered parton and were biased to select jets coming from the surface of the medium~\cite{renk} (geometrical bias). 

In the last couple of years, a step forward was made toward full jet reconstruction in heavy-ion collisions, enabled by the modern jet-finding techniques well suited for high-multiplicity environment.
In these proceedings, we report an overview of the results from full jet reconstruction in STAR~\cite{helen,jan,mateusz,elena}.

Unbiased full jet reconstruction allows a measurement of the energy of the hard scattering, even in the presence of the underlying heavy-ion collision.
Assuming that jets are a well calibrated pQCD probe (i.e. the predicted pQCD energy distribution agrees with reference measurements such as in p+p collisions), the inclusive jet production cross section should follow binary collision scaling from p+p
to Au+Au collisions. 
Unbiased reconstruction of jets should also allow the measurement of the softening fragmentation functions (FF) in Au+Au compared to baseline p+p jets of the same energy, as predicted by theoretical models~\cite{wiedemann}.

\section{Experimental techniques}
The STAR experiment is well suited for full jet reconstruction analysis thanks to the complete azimuthal coverage and pseudo-rapidity acceptance $|\eta|<1$ of the Time Projection Chamber (TPC) and the Barrel Electromagnetic Calorimeter (BEMC) that allow the reconstruction of charged particles and neutral energy, respectively.

The results presented in these proceedings are based on p+p year 2006 and central Au+Au year 2007 events. 

Two online triggers are used for the p+p data set: the High Tower (HT) and the Jet Patch (JP) trigger. The HT trigger requires, in addition to the minimum bias condition, one tower in the BEMC above a transverse-energy threshold $E_t$ of 5.4 GeV. In the JP trigger the total energy of a larger area of $\Delta \eta \times \Delta \phi=1 \times 1$ has to be above a threshold of 8 GeV.
The two triggers used for the Au+Au data set are the minimum bias trigger (MB) and the HT trigger in the BEMC. For Au+Au, the $E_t$ threshold for HT is 5.75 GeV. 

Jets are reconstructed using the ``$k_T$" and ``anti-$k_T$" recombination algorithms that are part of the FastJet package~\cite{fastjet,antikt} and are collinear and infrared safe. In the ``$k_T$'' algorithm, pairs of particles are clustered according to their spatial separation with a distance scale R (resolution parameter of the jet algorithm) and a weight given by the $p_T$ of the particles. In ``anti-$k_T$'', the weight is given by the inverse $p_T$ of the particles. Therefore, hard particles are clustered before soft particles. As a consequence, ``anti-$k_T$"  jets have a more circular shape and their radius is defined by the R parameter. 

A data-driven correction scheme for detector resolution and background effects was developed, as well as a first attempt to address the systematic uncertainties in the measurements~\cite{helen,mateusz,elena}.
\section{Background in jet analyses in Au+Au}
The background characterization is a challenge in full jet reconstruction because of the presence of the underlying event  in heavy-ion collisions. The $p_T$ of the reconstructed jet is conceptually defined in FastJet as:
$$
p_{T,rec}^{jet}=p_{T,true}^{jet} + \rho A \pm \sigma \sqrt A
$$
Fastjet provides an estimate of the background $p_T$ per unit area $\rho$, that is subtracted to get the jet component p$_{T,rec}^{jet}$.
The fluctuations of the background in a central Au+Au event are approximated by a Gaussian function. The width $\sigma$ is expected to scale with R and is of the order of 6-7 GeV in case of R=0.4. Non-statistical background fluctuations~\cite{ptcorrfluctu} remain to be included.
A resolution parameter $R=0.4$ suppresses the background without losing a significant fraction of the jet energy ($\sim$80$\%$ of the jet energy lies within R=0.4 for 20 GeV p+p jets~\cite{helen}).
 Upward fluctuations of uncorrelated background particles can be clustered into ``fake jets''. The fake jet spectrum is estimated using different procedures in the MB and HT Au+Au triggers.
 In the MB Au+Au data, the fake jet spectrum is estimated by removing the leading particles from the jets and running the jet-finding algorithms after randomizing in azimuth the remaining particles (tracks and towers).
The background jet spectrum in di-jet analysis (HT Au+Au data) is given by fake jets and jets originating from additional hard scatterings not associated with the HT trigger. The rate of background jets is estimated from the jet spectrum at $90^{\circ}$ in $\phi$ from the trigger jet.

The background fluctuations are corrected via an unfolding procedure after removing the background jet contribution from the measured jet spectrum, both in MB and HT Au+Au data sets.

\section{Jet measurements in p+p and d+Au}
\begin{figure}
\centering

\resizebox{0.45\textwidth}{!}{  \includegraphics{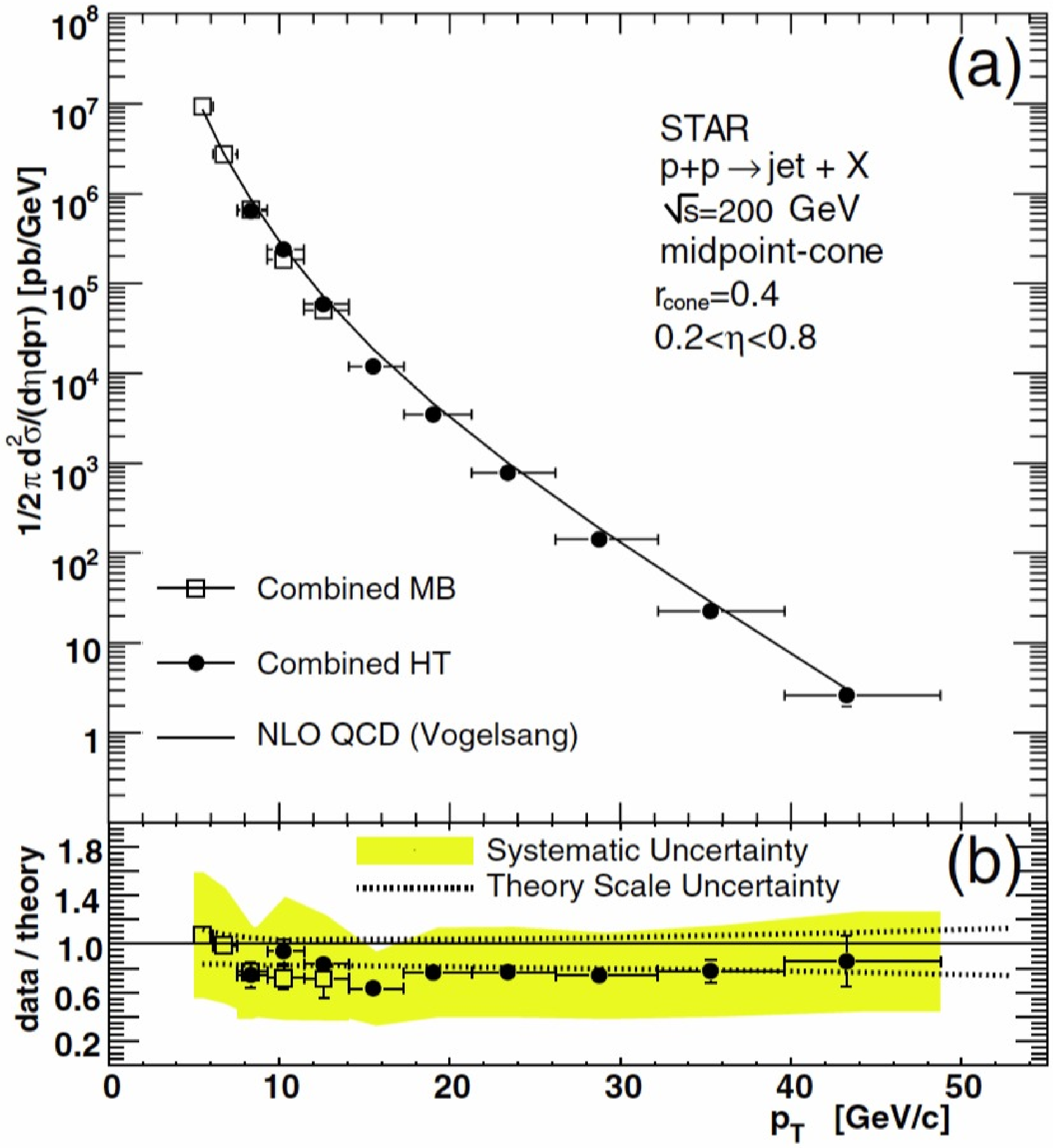}}
\resizebox{0.45\textwidth}{!}{  \includegraphics{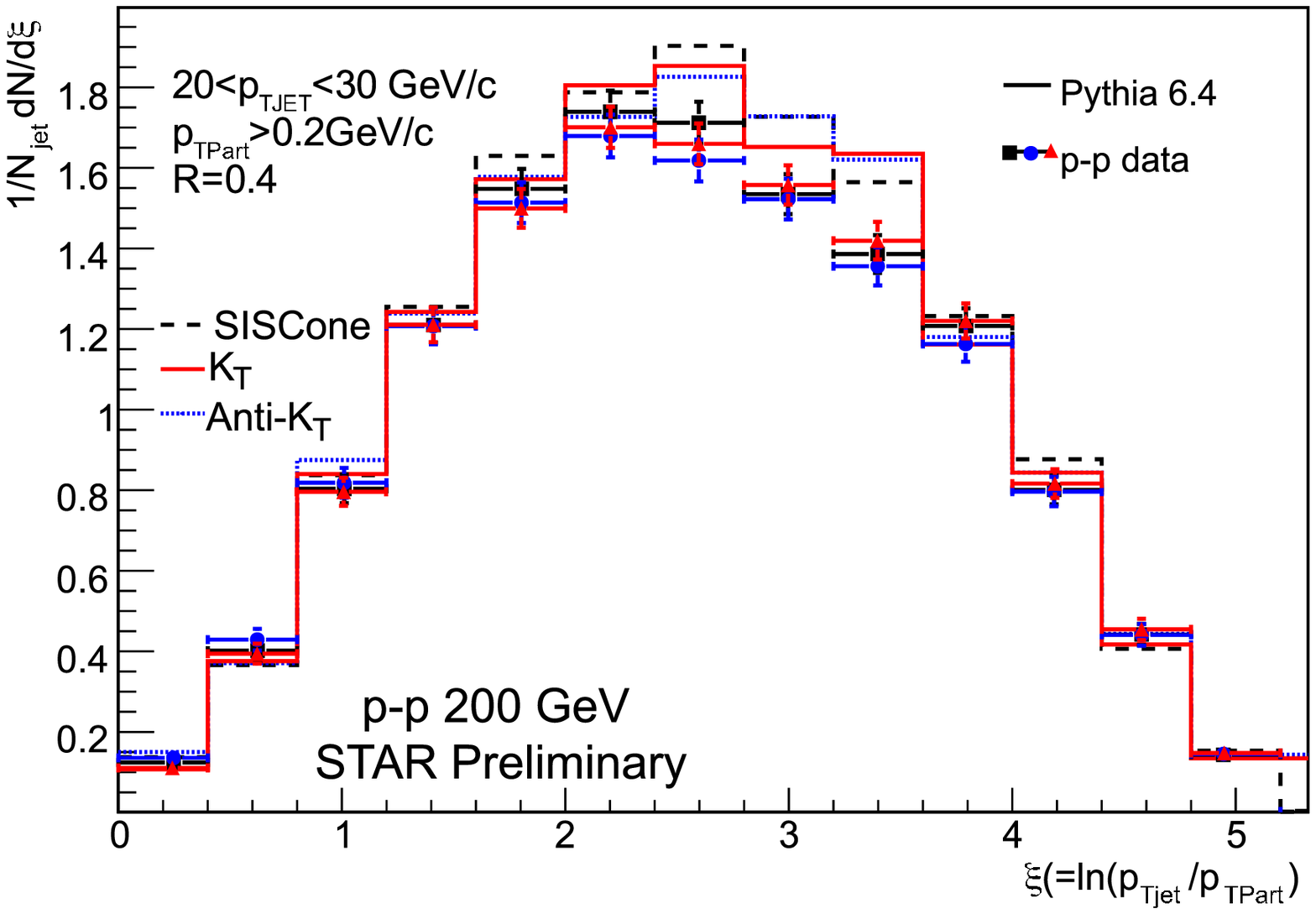}}

\caption{Left: (a) Inclusive differential jet cross sections for MB (open squares) and HT (filled circles) p+p data from the years 2003 and 2004. The curve shows a NLO calculation. (b) Ratio between data and theory. Right: $\xi$ distributions for jets reconstructed in p+p with $20<p_{T,rec}^{jet}<30$ GeV/c and R=0.4, compared to PYTHIA for three different jet finding algorithms, $k_T$ (red triangles), anti-$k_T$ (blue circles) and SISCone (black squares).}
\label{fig:ppxsec}

\end {figure}

The study of jet properties in p+p collisions is important to understand the production mechanism in QCD as well as the hadronization process at RHIC energies. In addition, jet  measurements in p+p and d+Au are needed as references to interpret the medium effects in Au+Au.
The very good agreement between the measured jet cross section in p+p and the NLO calculation over several orders of magnitude, as shown in Fig.~\ref{fig:ppxsec}~\cite{STARppxsec}, indicates that jets in p+p may be considered as a calibrated probe and can be used as a baseline for comparison to Au+Au results.
STAR also reported fragmentation functions (FF) in p+p~\cite{helen}. The FF as a function of $\xi=\ln(p_{T,rec}^{jet}/p_{T,hadr})$ for jets with reconstructed $p_T$ in the range 20 - 30 GeV are shown in Fig.~\ref{fig:ppxsec} (right) for R=0.4, for different jet finding algorithms applied to data and PYTHIA simulations. There is a reasonable agreement between data and PYTHIA. The FF are not corrected yet at the particle level, therefore the comparison is made by passing PYTHIA simulations through STAR's simulation and reconstruction algorithms.

\begin{figure}
\centering

\resizebox{0.65\textwidth}{!}{  \includegraphics{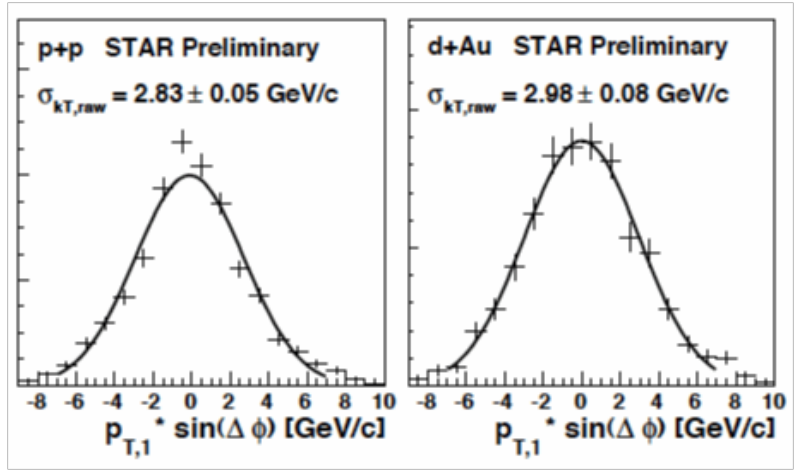}}

\caption{Uncorrected $k_{t,raw}$ for p+p (left) and d+Au (right) collisions, for $10<p_{T,2}<20$ GeV/c.}
\label{fig:jankt}
\end {figure}

Full jet reconstruction in d+Au allows exploration of possible cold nuclear matter effects, such as an additional $k_t$ broadening. In particular, STAR performed a measurement of  $k_{t,raw}=p_{T,1}\sin(\Delta\phi)$ via di-jets~\cite{jan}, with $p_{T,1}>p_{T,2}$. A study based on PYTHIA jets embedded in real d+Au events suggests that the contribution of the jet energy resolution and possible underlying background to the measured uncorrected $k_{t,raw}$ is small.  Fig.~\ref{fig:jankt} shows the $k_{t,raw}$ distributions for p+p and d+Au. The similar widths indicate that nuclear effects are rather small. 

\section{Inclusive jet measurements}\label{sec:inclu}
\begin{figure}
\centering
\resizebox{0.45\textwidth}{!}{  \includegraphics{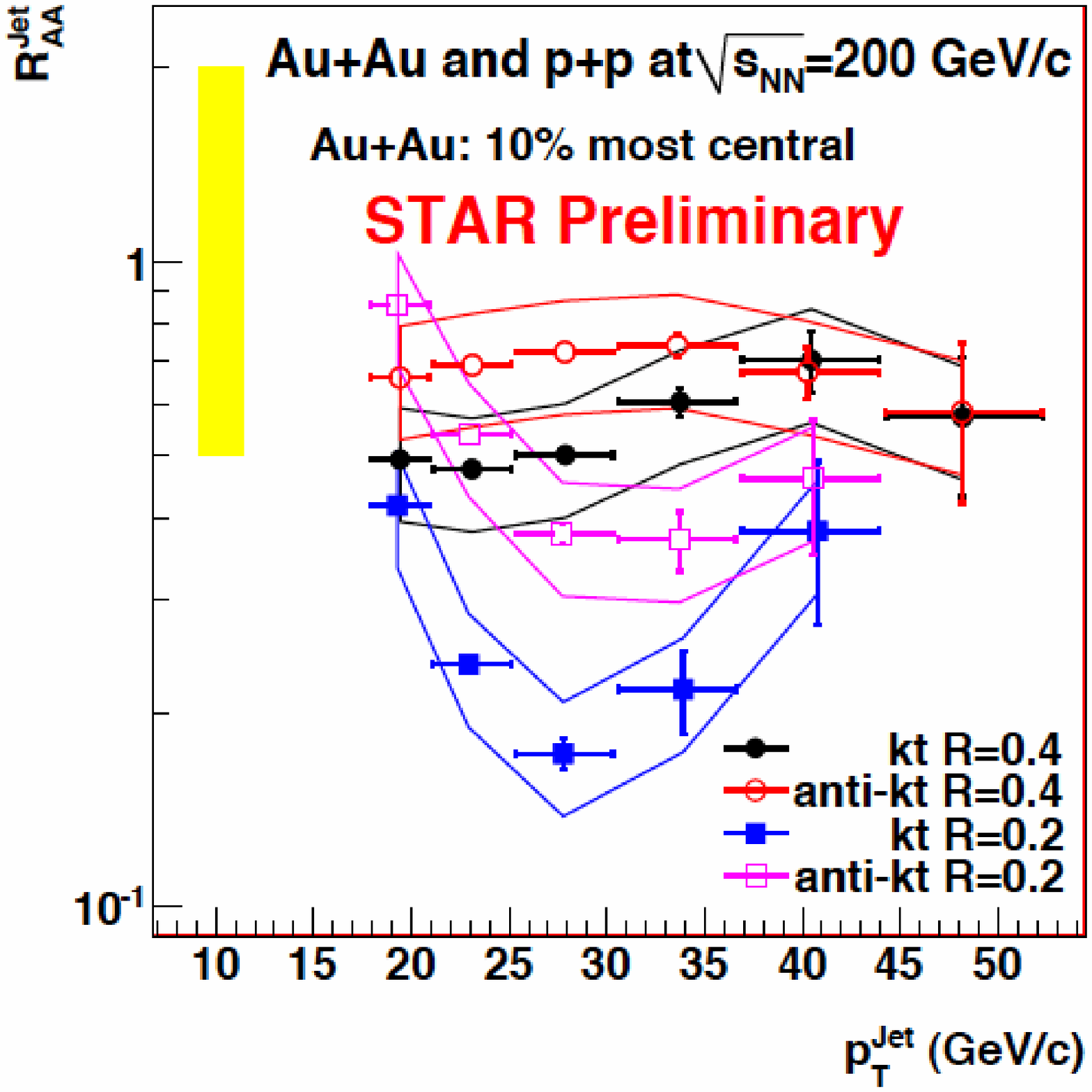}}
\resizebox{0.45\textwidth}{!}{  \includegraphics{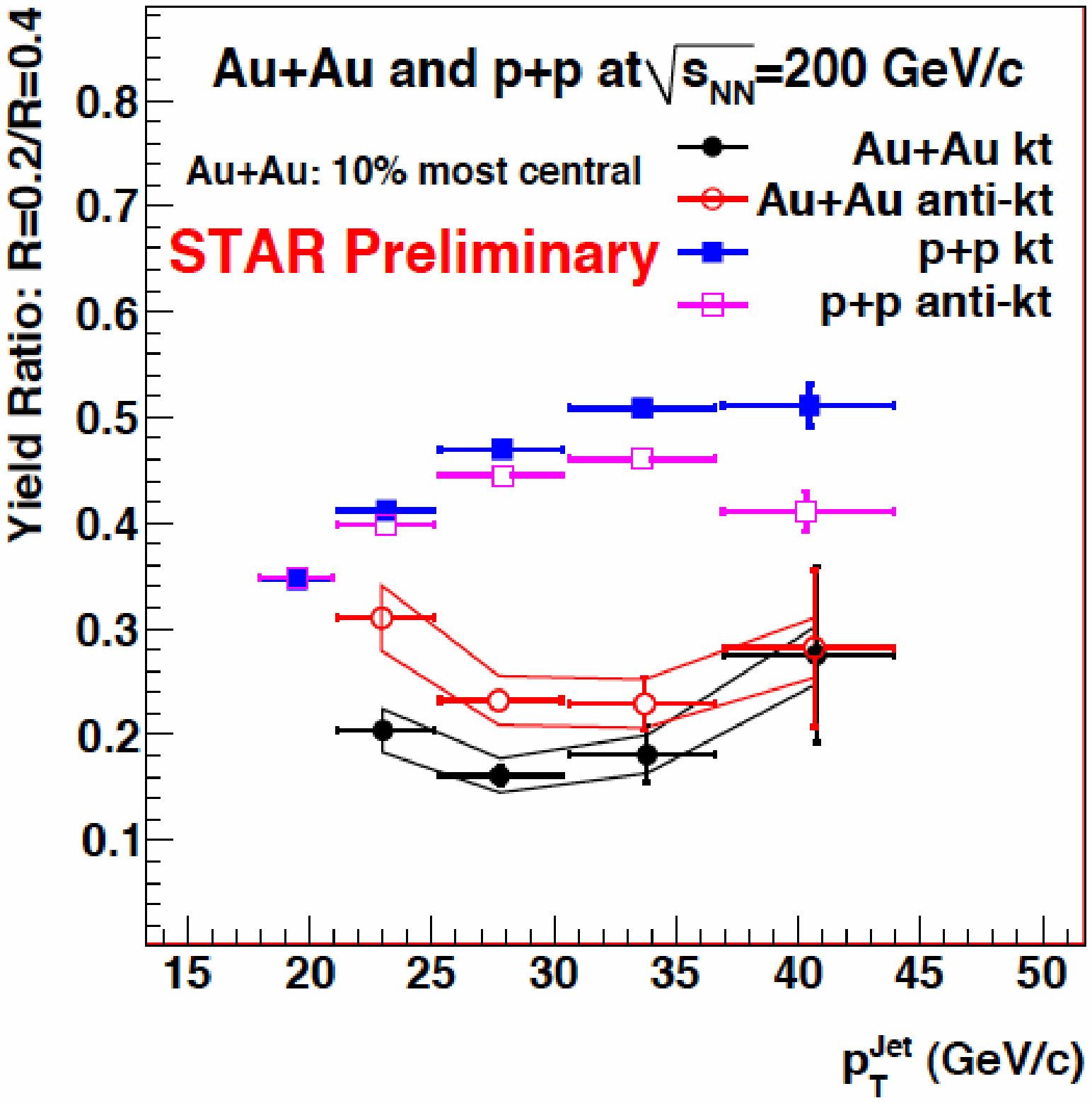}}

\caption{Left: $R_{AA}^{jet}$ for 0-10$\%$ most central MB Au+Au events. $k_T$ and anti-$k_T$ algorithms are used with R=0.4 and 0.2. The systematic uncertainty is shown as the vertical yellow band. Solid lines represent systematic uncertainties due to the unfolding of the background fluctuations. Right: Ratio of jet yields R=0.2/R=0.4, using both $k_T$ and anti-$k_T$, for each collision system.}

\label{fig:AuAuspectra}
\end {figure}

The $R_{AA}^{jet}$, reported in Fig.~\ref{fig:AuAuspectra} (left)~\cite{mateusz}, is the ratio of the jet yield in  0-10$\%$ most central MB Au+Au events to the jet yield in p+p scaled by the number of binary collisions. $k_T$ and anti-$k_T$ algorithms were used with two choices of R parameter, 0.4 and 0.2. The kinematic range extends up to about 50 GeV/c, significantly increased as compared to inclusive single particle measurements.
The measurements are corrected for detector resolution and background fluctuations. The systematic uncertainty in the jet energy scale is dominated by the BEMC calibration ($5\%$), the unobserved neutral energy ($3\%$)
and the momentum resolution of the charged particles ($2\%$). 

In the case of an unbiased jet reconstruction,  $R_{AA}^{jet}$ is expected to be close to unity because the jet production is a hard process and should scale as $N_{bin}$. It can be noticed that the  $R_{AA}^{jet}$ for R=0.4 recovers more jets compared to single particle $R_{AA}$ and R=0.2. However, even though the systematic uncertainties are quite large, there are indications that unbiased jet reconstruction is not achieved even for R=0.4 because  $R_{AA}^{jet}$ does not approach 1 at large $p_{T}^{jet}$. Possible explanations could be an overall absorption or that significant broadening of the jet energy profile beyond R=0.4 occurs in Au+Au, which would cause an underestimation of the reconstructed jet energy in Au+Au relative to p+p.
Measurements of the jet energy profile could help to distinguish between the two phenomena. Fig.~\ref{fig:AuAuspectra} (right) shows the ratio of the jet yields for R=0.2 over R=0.4 for p+p and Au+Au. The ratio increases with the jet $p_T$ in p+p, consistent with PYTHIA calculations which predict that jets become more collimated as their energy increases. The ratio is significantly suppressed for Au+Au, indicating a broadening of the jet energy profile going from R=0.2 to R=0.4. Assuming a continuous evolution, it is expected that the broadening mechanism extends beyond R=0.4.

\section{Di-jet analysis}
\begin{figure}
\centering
\resizebox{0.48\textwidth}{!}{  \includegraphics{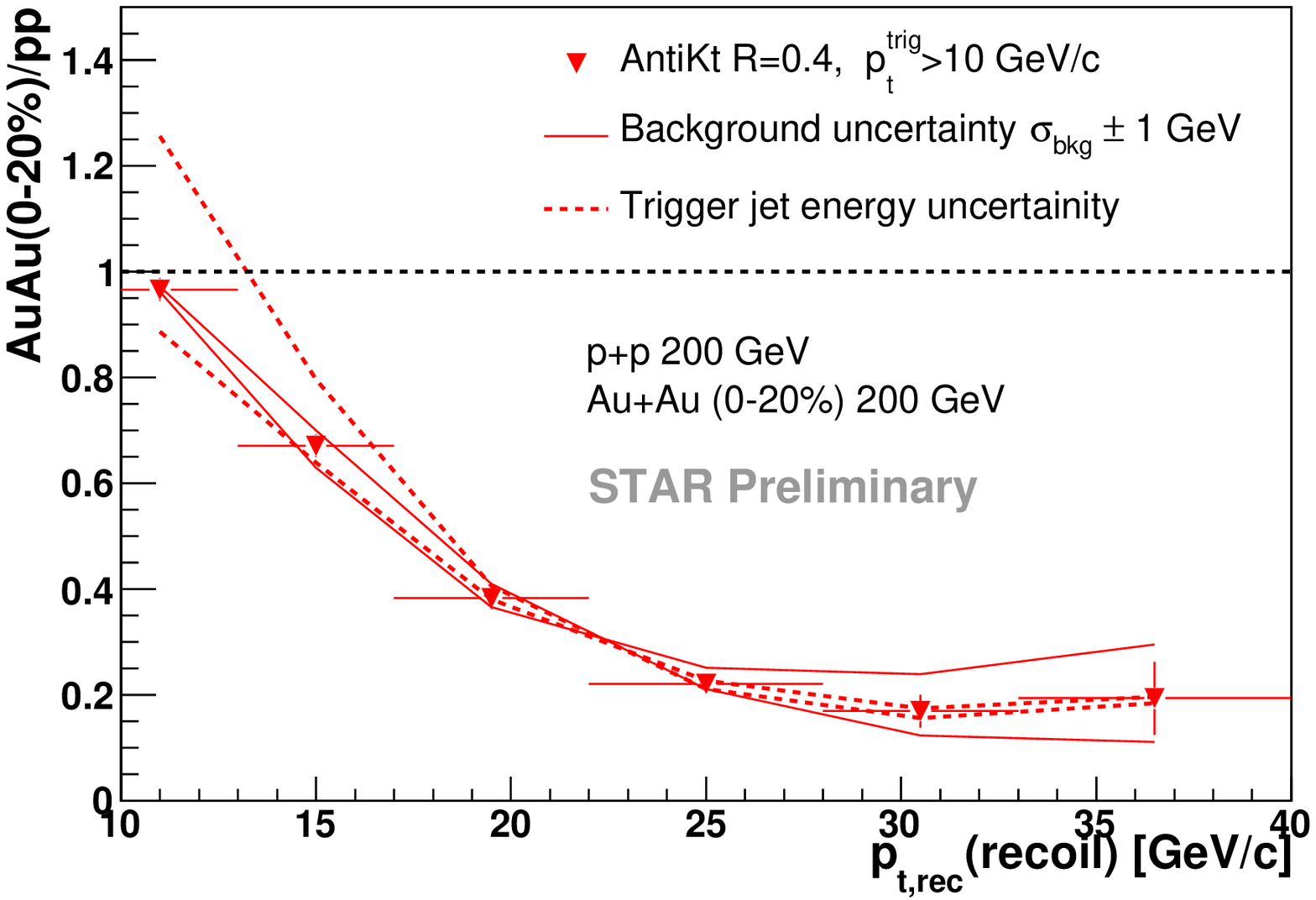}}
\resizebox{0.46\textwidth}{!}{  \includegraphics{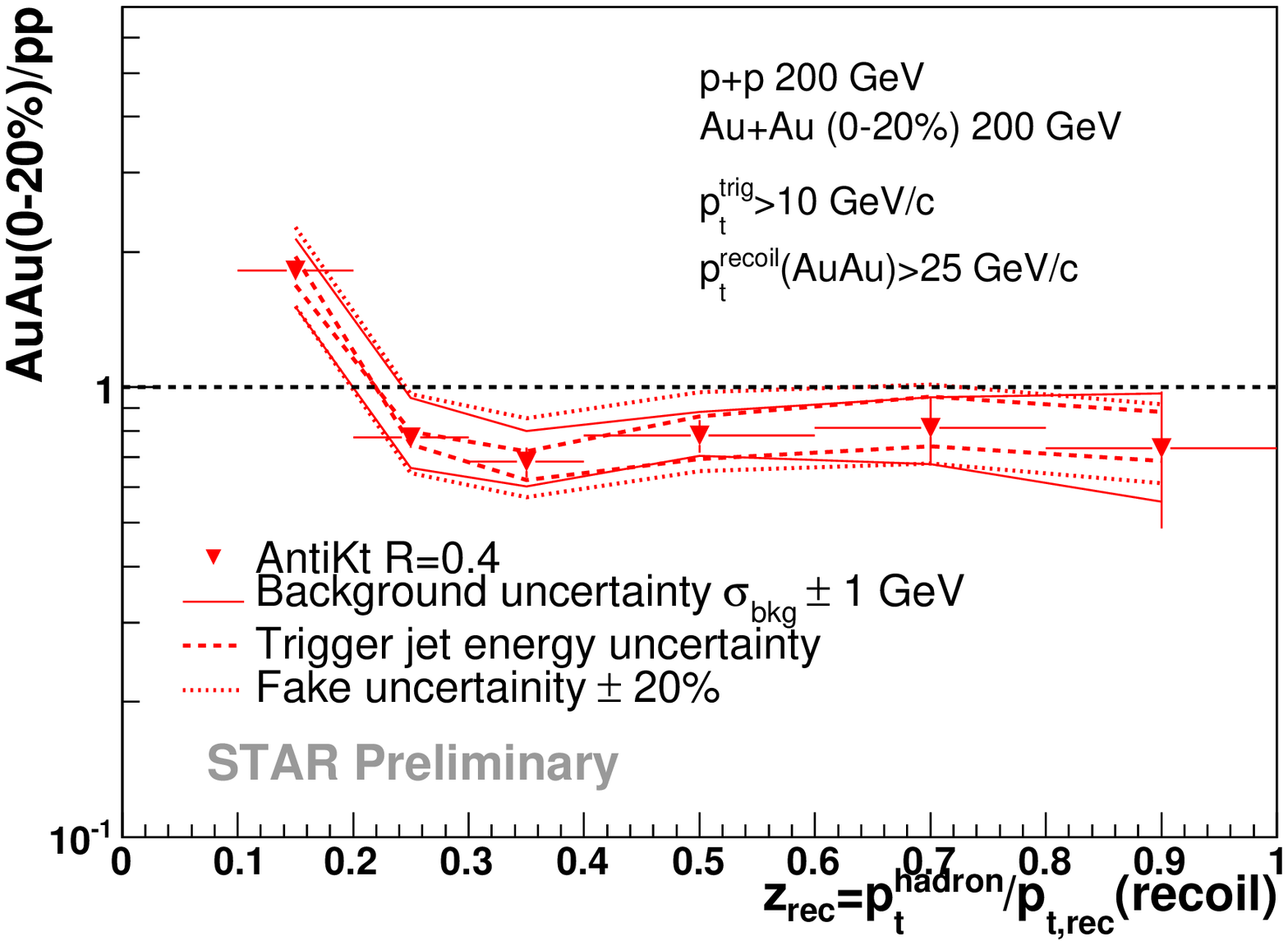}}

\caption{Left: ratio of $p_t$ spectra of recoil jets in Au+Au (corrected for fake jets and background fluctuations) to p+p. The curves indicate the systematic uncertainties in the estimation of the background fluctuations (solid) and of the $p_{t}$ of the trigger jet assuming a jet-$p_t$ resolution of 25$\%$ (dotted)~\cite{helen}. Right: ratio of the $z$ distributions measured in 0-20$\%$ central Au+Au events to p+p collisions for $p_{t,rec}^{recoil}(AuAu)>25$ GeV/c. The curves indicate the systematic uncertainties in the estimation of the background fluctuations (solid) and of the trigger jet energy (dotted). The uncertainty due to the estimation of fake jets (small dotted curves) was assumed to only affect the normalization of the $z$ distributions.}
\label{fig:ppAuAuDiJets}
\end{figure}

STAR performed a measurement of di-jets in HT p+p and 0-20$\%$ HT Au+Au events~\cite{elena}, using the anti-$k_T$ algorithm with R=0.4. The analysis searched for a ``trigger'' jet matching the online triggered tower in the BEMC and a ``recoil'' jet on the away side of the trigger jet (i.e. $\Delta \phi \sim \pi$).
For the trigger jet a $p_t^{cut}=2$ GeV/c was applied to tracks and towers in order to have the same energy scale in p+p and Au+Au. The $p_t^{cut}$ introduces a strong bias in the jet population in Au+Au, causing the trigger jets to more likely come from the surface of the medium. In order to avoid a similar surface bias on the recoil jets, the latter were reconstructed with a minimal  $p_t^{cut}=0.15$ GeV/c given by the TPC acceptance. The requirement of $p_{t,rec}^{jet}(trigger)>10$ GeV/c for the reconstructed trigger jet minimizes the contribution of ``fake'' jets. 
The results of this analysis are not corrected for the difference in tracking efficiency between p+p and Au+Au. An absolute correction to get the parton energy will be performed in the future.
However, we applied a data-driven unfolding of the background fluctuations from the Au+Au di-jet spectra and FF of recoil jets to directly compare p+p and Au+Au. 
The ratio of recoil jet spectra normalized to the number of trigger jets in Au+Au over p+p is shown in Fig.~\ref{fig:ppAuAuDiJets} (left). A significant suppression of recoil jets is observed in Au+Au relative to p+p, in contrast with the expected ratio of 1 for unbiased jet reconstruction. This could be explained by a broadening of the jet structure in an area beyond R=0.4 (see Sec.~\ref{sec:inclu}), resulting an underestimation of the jet energy within R=0.4 in Au+Au relative to p+p. Such an effect can be attributed to the trigger bias that forces the recoil jets to have large in-medium path-length and hence suffer a maximum energy loss.

STAR also measured the FF of recoil jets in p+p and Au+Au. 
The FF are measured for charged hadrons within  R=0.7 around the jet axis, whereas the reconstructed jet energy is defined for R=0.4. This method includes additional soft fragmentation particles.
The charged particles included in the FF in Au+Au contain the jet fragments plus background particles., where the latter must be subtracted in order to measure the jet fragmentation. The background component is estimated on an event-by-event basis from the charged particle spectrum in the area outside the two jets with the highest reconstructed energy. 
The FF are measured in Au+Au for $p_{t,rec}^{recoil}(AuAu)>25$ GeV/c. The p+p jets corresponding to  $p_{t,rec}^{recoil}(AuAu)>25$ GeV/c are selected taking the background fluctuations into account, assuming a Gaussian parameterization with $\sigma=6$ GeV. The mean jet $p_t$ of the selected p+p jets is $p_{t,rec}^{recoil}(pp)\simeq 25$ GeV/c. 
The ratio of the $z=p_{t,hadr}/p_{t,jet}^{recoil}$ distributions of Au+Au to p+p is shown in Fig.~\ref{fig:ppAuAuDiJets} (right).

No significant modification of the fragmentation functions is observed for recoil jets in Au+Au relative to p+p, for $z > 0.2$. This is in contrast with the theoretical expectation of a softening of the FF for an unbiased jet population  and with the measured suppression of high-p$_t$ hadrons at RHIC ($R_{AA} \sim 0.2$~\cite{Raa}), both of which are related to parton energy loss in the medium. The low-$z$ part of the fragmentation function ($z < 0.1$) is still under investigation because it might be affected by a large uncertainty due to background subtraction.

\section{Summary}
We presented in these proceedings an overview of the STAR results on jet studies.
Full jet reconstruction extends the kinematic reach of studies of parton energy loss mechanisms in the dense medium at RHIC.
Jet measurements in p+p show that full reconstructed jets are well calibrated tools to probe the medium.
Preliminary results from d+Au suggest that there are no major cold nuclear matter effects which would cause additional $k_t$ broadening in di-jet measurements.

A suppression of $R_{AA}^{jet}$ was observed  in inclusive jet measurements for $p_{t,jet}>20$ GeV/c, suggesting that the jet energy is not fully recovered within R=0.4 and R=0.2. This is consistent with the observed decrease of jets with R=0.2 relative to R=0.4  in Au+Au with respect to p+p, indicating a broadening of the jet profile as observed in going from R=0.2 to R=0.4.

In di-jet analysis, the online trigger maximizes the in-medium path of the recoil jets making them more susceptible to energy loss. Given this selection bias, a suppression of the recoil jets in Au+Au relative to p+p was observed. This suppression can be explained with a scenario where the jet is broadened and its energy is not fully recovered in $R=0.4$ with respect to p+p with the current jet-finding algorithms, causing a shift of the recoil spectrum in Au+Au towards smaller energies.
The absence of a strong modification of the FF in Au+Au with respect to p+p could be due to an underestimation of the jet energy within R=0.4 owing to the preceding jet broadening picture.

Future measurements will attempt to determine wether or not jet broadening occurs beyond R=0.4, as suggested by the $R_{AA}^{jet}$ and di-jet spectra measurements. The jet-energy profile will also be investigated in order to improve our understanding of the mechanism of jet broadening in the medium.
Further systematic studies, including a detailed background characterization, are being conducted in order to access the maximum possible kinematic range of the hard scattering.

\section{Acknowledgments}
The author wishes to thank the Bulldog facility at the Yale University.

\end{document}